\documentclass[english,final,3p,times]{elsarticle}
\usepackage[T1]{fontenc}
\usepackage[latin9]{inputenc}
\usepackage{array}
\usepackage{amstext}
\usepackage{color}

\makeatletter

\usepackage{graphicx}

\usepackage{amssymb}

\usepackage{lineno}
\usepackage{mathrsfs}

\usepackage{multicol}

\journal{Computer Physics Communications}

\makeatother

\usepackage{babel}

\begin{document}
\begin{frontmatter}

\title{Multi-GPU Accelerated Multi-Spin Monte Carlo Simulations of the 2D Ising Model\tnoteref{label_title}}

\tnotetext[label_title]{Source code of our implementations for GPU clusters will be published on \texttt{http://www.tobiaspreis.de} after acceptance. In addition, the code can be downloaded from the Google Code project \emph{multigpu-ising}.}

\author[uni]{Benjamin Block\corref{corr}}
\ead{lhyanor@gmail.com}
\author[uni]{Peter Virnau}
\ead{virnau@uni-mainz.de}
\author[uni]{Tobias Preis}
\ead{mail@tobiaspreis.de}
\cortext[corr]{Corresponding author} 
\address[uni]{Department of Physics, Mathematics and Computer Science,\\ Johannes Gutenberg University Mainz -- Staudingerweg 7, D-55128 Mainz, Germany}

\begin{abstract}
A modern graphics processing unit (GPU) is able to perform massively parallel scientific computations at low cost. We extend our implementation of the checkerboard algorithm for the two dimensional Ising model [T.~Preis {\it et al.}, J.~Comp.~Phys.~{\bf 228}, 4468 (2009)] in order to overcome the memory limitations of a single GPU which enables us to simulate significantly larger systems. Using multi-spin coding techniques, we are able to accelerate simulations on a single GPU by factors up to $35$ compared to an optimized single Central Processor Unit (CPU) core implementation which employs multi-spin coding. By combining the Compute Unified Device Architecture (CUDA) with the Message Parsing Interface (MPI) on the CPU level, a single Ising lattice can be updated by a cluster of GPUs in parallel. For large systems, the computation time scales nearly linearly with the number of GPUs used. As proof of concept we reproduce the critical temperature of the 2D Ising model using finite size scaling techniques.
\end{abstract}


\end{frontmatter}

\section{Introduction}

Various scientific disciplines profited by GPU computing in recent years and are reporting impressive speedup factors in comparison to single Central Processor Unit (CPU) core implementations. GPU stands for Graphics Processing Units which are high-performance many-core processors that can be used to accelerate a wide range of applications. In the meantime, significant savings of computing time have been reported by a huge variety of fields: GPU acceleration can be used in astronomy \cite{Ford09} and radio astronomy \cite{Harris08}. Soft tissue simulation \cite{Tay09}, algorithms for image registration \cite{Gu09a}, dose calculation \cite{Gu09b}, volume reconstruction from x-ray images \cite{Gro09}, and the optimization of intensity-modulated radiation therapy plans \cite{Men09} are examples for the numerous applications in medicine. Furthermore, DNA sequence alignment \cite{Trap09}, molecular dynamics simulations \cite{And08,Fried09,Mee08}, quantum chemistry \cite{Ufim08}, multipole calculations \cite{Gume08}, density functional calculations \cite{Gen09,Yas08}, air pollution modeling \cite{Mol10}, time series analysis focused on financial markets \cite{Pre09b,Pre08}, and Monte Carlo simulations \cite{Yin09,Bad09,Mere09,Preis09} benefited from GPU computing. For many applications, the accuracy can be comparable to that of a double-precision CPU implementation, such as in \cite{Harv09}---the latest generation of GPUs support not only single precision but also double precision floating point operations. The adaption of many computational methods is still in progress, e.g. the analysis of switching processes in financial markets \cite{Pre09c,Pre10a}. 
Unfortunately, not all algorithms can be ported efficiently onto a GPU architecture. Particularly, serial algorithms are not suited for GPU computing (for an example see e.g. \cite{Rei10}). 

Another crucial limitation is the lack of scalability as current programs typically utilize only single GPUs. As graphics processing hardware is targeted at a broad consumer market---the games industry---, graphic cards can be produced at low cost. On the other hand, to keep production costs low, the global memory is not upgradable and typically limited to $1$ GB for consumer cards and $4$ GB for Tesla GPUs. Using a recent consumer graphics card, we accelerated Monte Carlo simulations of the Ising model \cite{Preis09}. In \cite{Preis09}, a 2D square spin lattice of dimension up to $1024^2$ spins could be processed on a consumer GPU. The Ising model as a standard model of statistical physics provides a simple microscopic description of ferromagnetism \cite{Isi25}. It was introduced to explain the ferromagnetic phase transition from the paramagnetic phase at high temperatures to the ferromagnetic phase below the Curie temperature $T_C$. A large variety of techniques and methods in statistical physics have originally been formulated for the Ising model and were generalized and adapted to related models and problems \cite{Bin01}. Due to its simplicity, which can be embodied by the possibility to use trivial parallelization approaches \cite{Ito93}, the two dimensional Ising model is well suited as a benchmark model since its properties are well studied \cite{Lan05,Sta87,Sta92} and many physical systems belong to the same universality class. The Ising model on a two dimensional square lattice with no magnetic field was analytically solved by Lars Onsager in 1944 \cite{Ons44}. The critical temperature at which a second order phase transition between an ordered and a disordered phase occurs can be determined analytically for the two dimensional model ($T_C\approx2.269185$ \cite{Ons44}).

Here we show how lattice sizes can be extended up to $100,000^2$ spins on one GPU device with 4 GB of global memory using a memory optimized encoding of the spins---one bit per spin. This number of spins turns out to be a hard limitation on a single device, since for larger system sizes, spin data would have to be transferred between device and host memory. Such a memory transfer would effectively rule out all performance benefits of a GPU implementation. Using a multi-spin coding scheme \cite{Wansleben84,Zorn81,Ito88,Ito90}, computation to memory access ratio can be improved, resulting in a dramatically faster GPU performance. 

We show that an extension of this approach can be used successfully to handle Monte Carlo simulations of the Ising model in a multi-GPU environment---GPU clusters. The scalability of this implementation is ensured by splitting the lattice into quadratic sublattices, and by placing them into the memory of different GPUs. Thus, each GPU can perform the calculation of one sublattice in its memory and pass the information about its borders on to its neighboring GPUs. Similar approaches have been used, e.g., for the calculation of density functionals \cite{Gen09}.

This paper is organized as follows. In a brief overview in Sect.~\ref{sec_gpu_device}, key facts of the GPU architecture are provided in order to clarify implementation constraints for the following sections. Sect.~\ref{sec_ising} provides a survey of model definition and finite size scaling techniques used as proof of concept. In Sect.~\ref{sec_cpu_implementation} and Sect.~\ref{sec_gpu_implementation}, we describe details of the reference CPU implementation and our single GPU approach based on multi-spin coding. The multi-GPU accelerated Monte Carlo simulation of the 2D Ising model is covered in Sect.~\ref{sec_multi_gpu}. To overcome the memory limitations of a single GPU with such a multi-GPU approach is of crucial importance as GPU clusters are currently set up in supercomputing facilities. Our conclusions are summarized in Sect.~\ref{sec_conclusion}.

\section{GPU Device architecture}\label{sec_gpu_device}

Simulating the Ising model at large system sizes requires a lot of processing performance. Physical and engineering obstacles in microprocessor design have resulted in flat performance growth for traditional single-core microprocessors. On the other hand, graphics hardware has become highly programmable, the fixed function pipelines have been replaced by programmable shader units that can perform the same operation on many sets of data in parallel. For a comprehensive overview of recent developments in computer graphics, especially programmable shader techniques, see \cite{Moe08}. With new, more flexible programming interfaces, these units can be utilized to perform general purpose computing in fields other than computer graphics.

For our GPU implementation, we use the Compute Unified Device Architecture (CUDA) released by NVIDIA for their recent graphics accelerator boards. The latest stable release at the time of writing is CUDA 2.3 \cite{Nvi09a}. Recently, other Application Programming Interfaces (APIs) for General Purpose computing on GPUs (GPGPU) became available, see e.g. OpenCL \cite{Nvi09b}. Additionally, efforts have been made to establish high-level programming environments \cite{Mess08} as well as the integration into existing compilers.

We use a NVIDIA Tesla C1060 as our CUDA enabled device, which offers $4$ GB of GDDR3 global memory, see Table~\ref{tab_device}. This memory can store a multi-spin coded spin field of $100,000^2$ spins on one GPU. The reference CPU used in our tests is the Intel Xeon X5560 at a clock rate of $2.80$ GHz and $8192$ kB cache. The purpose of the CPU implementation is to have a fast and fair non-parallel reference implementation, not to benchmark the Intel CPU. Therefore, only one core of the CPU is used (without Hyper-Threading Technology).

CUDA implements a Single Instruction Multiple Thread (SIMT) approach. It is capable of running the same code in parallel, processed in a {}``grid''. A grid is a number of blocks which in turn contain a defined number of threads. It extends the C language by the invocation of {}``kernels'' that run in parallel in such a grid on the GPU:
\begin{verbatim}
cuda_kernel<<<gridDim, blockDim>>>(data);
\end{verbatim}
The variable {\it gridDim} defines the number of blocks that run in parallel, and {\it blockDim} specifies the number of threads that run in each block. Threads in each block share a certain amount of {}``shared memory'' which can be accessed roughly one order of magnitude faster than data in the global GPU memory. The Tesla C1060 is capable of processing a maximum of $512$ threads per block. Kernels are executed on the actual hardware in units of {}``warps'', where each warp executes one common instruction at a time. 

With a larger number of threads in a block, memory latencies can be hidden more effectively. However, the hiding of memory latencies only results in better performance, if the number of registers used by a single thread is sufficiently small. To optimize execution time for a kernel, a grid size should be used that allows for a maximum {}``occupancy''. The occupancy is the ratio of active warps to the maximum number of warps supported on a multiprocessor of the GPU. A multiprocessor contains amongst others eight scalar processor cores, a multi-threaded instruction unit, and shared memory. This ratio is a helpful number to determine how efficient the kernel will be on the GPU.

\begin{table}
\caption{Key facts and properties of the GPU\cite{Nvi09c}.}
\label{tab_device}
\begin{center}
\begin{tabular}{lll}
\hline\noalign{\smallskip}
  & Tesla C1060 \\
\noalign{\smallskip}\hline\noalign{\smallskip}
  Global video memory & $4096$ MB\\
  Streaming processor cores & $240$\\
  Shared memory per block & $16$ KB\\
  Processor clock & $1.30$ GHz\\
  Memory clock & $800$ MHz\\
  Maximal power consumption & $187.8$ W\\
\noalign{\smallskip}\hline
\end{tabular}
\end{center}
\end{table}

For the multi-GPU implementation, each CPU core runs a separate process and controls one of the available GPUs. Communication is established via the Message Passing Interface (MPI). Communication is needed frequently (see Sect.~\ref{sec_multi_gpu}) which leads to a bottleneck for small systems, but is ruled out by the benefit of more available GPU cores for larger systems. 

\section{The two-dimensional Ising model}\label{sec_ising}

The Ising model is formulated on a two-dimensional square lattice, where on each lattice site a spin $S_i$ with a value of either $-1$ or $1$ is located. The interaction of the spins is given by the Hamiltonian
\begin{equation}
\displaystyle \mathscr{H} = - J \sum_{\langle{} i,j\rangle{}} S_i S_j -H \sum_i S_i
\end{equation}
where $H$ denotes an external magnetic field, which we will set to zero here. The lattice is updated according to the Metropolis criterion \cite{Met53}. For each step, the energy difference $\Delta \mathscr{H} = \mathscr{H}_{a} - \mathscr{H}_b$ between two subsequent states $a$ and $b$ is calculated. The probability for the step to be accepted is given by $W_{a\rightarrow b} = \exp (-\Delta \mathscr{H} / k_B T)$ if $\Delta \mathscr{H} > 0$ and $W_{a\rightarrow b} = 1$ if $\Delta \mathscr{H} \leq 0$. Since only discrete values for this factor are possible, they should be pre-calculated on the CPU for each temperature and transferred to the GPU when the kernels are invoked.

To make efficient use of the GPU device structure, a parallelizable spin-update scheme has to be utilized. The ratio between memory latency and processing time on graphics cards is very large \cite{Nvi09a}. Thus, GPU cores can perform hundreds of instructions in the time of a single access to the global memory. By highly parallel processing, memory access latencies can be hidden effectively, and large acceleration factors achieved.

Parallel spin updates of the Ising model can only be done for non-interacting domains. The approach that each spin only interacts with its four nearest neighbors makes a checkerboard update feasible \cite{Preis09}. The lattice update is divided into two update steps {\it A} and {\it B}. In step {\it A}, only the spins residing on a black site are updated since they are not interacting with each other. In step {\it B}, the spins on white lattice sites are updated. It is essential that update step {\it B} is started after all updates of step {\it A} are finished. Please note, that other methods for the spin updating process are also available, e.g. diverse cluster algorithms \cite{Swe87,Wol89}, perform particularly well close to the critical point. However, the systematic scheme of the checkerboard algorithm is most suitable for the GPU architecture realizing non-interacting domains where the Monte Carlo moves are performed in parallel.

In order to test the correctness of the implementation, we determine the critical temperature of the Ising spin system. We use finite size scaling and calculate the Binder cumulant \cite{Bin81,Lan05}
\begin{equation}
\displaystyle U_4\left(T\right)=1-\frac{\langle M(T)^4 \rangle}{3\langle M(T)^2 \rangle^2}
\end{equation}
with $M$ denoting the magnetization of a configuration at temperature $T$ and $\langle \ldots \rangle$ denoting the thermal average. Near a critical point, finite size scaling theory predicts the free energy and derived quantities like the magnetization to be a function of linear dimension $L$ over correlation length $\xi \simeq (T-T_C)^{-\nu}$. Therefore, moment ratios of the magnetization like, e.g. the Binder cumulant $U_4$, become independent of system size $N=n^2$ at the critical temperature $T_C$. To test our implementation, we perform several simulations close to the critical point for various linear dimensions $n$ of the simulation box and determine $U_4$.

\section{Optimized reference CPU implementation}\label{sec_cpu_implementation}

For our optimized CPU reference implementation, we focus on a single spin-flip approach which performs well for large lattice sizes. Multi-spin coding refers to all techniques that store and process multiple spins in one unit of computer memory. In CPU implementations, update schemes have been developed that allow to process more than one spin with a single operation \cite{Wansleben84,Zorn81,Ito88,Ito90}. We use a scheme which encodes $32$ spins into one $32$-bit integer in a linear fashion. The $32$-bit type is chosen since register operations of current hardware perform fastest on this data type. The key ingredient for an efficient update algorithm of these $32$-bit patterns is to use precomputed bit patterns that encode the evaluations of the flip condition expression
\begin{equation}
\displaystyle r <  \exp (-\Delta \mathscr{H} / k_B T)
\end{equation}
for every single spin bit---the variate $r$ is an independent and identically-distributed random number in $[0,1)$. Since there are only two possible energy differences $\Delta \mathscr{H}$ with $\Delta \mathscr{H}>0$, two Boolean arrays can encode the information of an evaluation of the flip condition.  For reasonable results, N.~Ito \cite{Ito88} suggested to use a pool of $2^{22}$ to $2^{24}$ Boltzmann patterns. 

We call the two Boolean arrays exp4 and exp8. Our encoding is chosen to store a 1 into exp4 if $\exp(-8J/k_BT) < r < \exp(-4J/k_BT)$ and a 0 if not, and a 1 into exp8 if $r < \exp(-8J/k_BT)$ and a 0 if not.

For every spin update, the Monte Carlo simulation will draw a random address in this pool, and return the bit patterns at this address. To process a $32$-bit spin pattern $s_0$, neighbor patterns $s_1, s_2, s_3, s_4$ have to be prepared, that contain the neighbors of the $i$th spin at their $i$th digit.

%

Since the calculation is the same for every bit, it is convenient to look at just one bit.
The first step is to transform the spin variables into \emph{energy variables} $i_n$ to get rid of the dependence of the initial state of $s_0$.

\begin{equation}
i_n = s_0 \text{\textasciicircum} s_n, \forall n \in \{1,2,3,4\}
\end{equation}
where \textasciicircum{} denotes an exclusive-or operation. Because of the special encoding of the Boltzmann patterns, the acceptance condition for each spin can be expressed in a simple way:

\begin{equation}
i_1 + i_2 + i_3 + i_4 + 2 \cdot \text{exp8}_s + \text{exp4}_s \geq 2
\end{equation}
where exp8$_s$ and exp4$_s$ denote the $s$th Boltzmann patterns that encode the spin flip condition, and $s$ is a random position in the pool. It is possible to evaluate this expression for all 32 bits at once by applying a sequence of bitwise Boolean operations \cite{Ito88}.


Since parallel updates are only allowed on non-interacting domains, an additional bit mask has to be applied to the update pattern, that only allows to flip every second spin at once. Details of the implementation can be found in \cite{Blo10}.

\section{Single-GPU implementation}\label{sec_gpu_implementation}

\subsection{Problems arising from a straightforward porting}

On graphics cards, memory access is very costly compared to operations on registers. The great advantage of multi-spin coding is that only one memory access is needed to obtain several spins at once. The CPU implementation could be ported to GPU with a kernel that uses less than $16$ registers. This allows an optimal occupancy of the GPU up to a maximum block size of $512$ threads. Even though the update scheme presented in Sect.~\ref{sec_cpu_implementation} performs vastly faster on the CPU compared to an implementation with integer representations of each spin, a straightforward GPU port of this scheme is not optimal.

The reason for the poor performance is that parallel threads in one warp have to access global memory in a random fashion which is very costly. The execution speed can be improved by drawing only one random position per block, and let all the threads in this block read the patterns linearly, starting from the drawn starting position. This however, reduces the quality of the flip patterns---this could in principle be compensated by using a significantly larger pool of random numbers. Another option is to calculate the spin flip patterns on the fly using a random number generator on the GPU (see \ref{sec_random_number}) instead of looking them up from the global memory. It turns out however, that the sophisticated update scheme has no benefit here, anymore. The performance of these implementations is compared in Table \ref{tab_comparing}. It should be emphasized that the quality of random numbers differs between the implementations. In the next section, we present another update scheme that works well on the GPU, which prevents pitfalls with the quality of the random numbers.

\subsection{Extraction into shared memory}\label{sec_shared_memory}

\begin{figure}[htb]
\includegraphics[width=\textwidth]{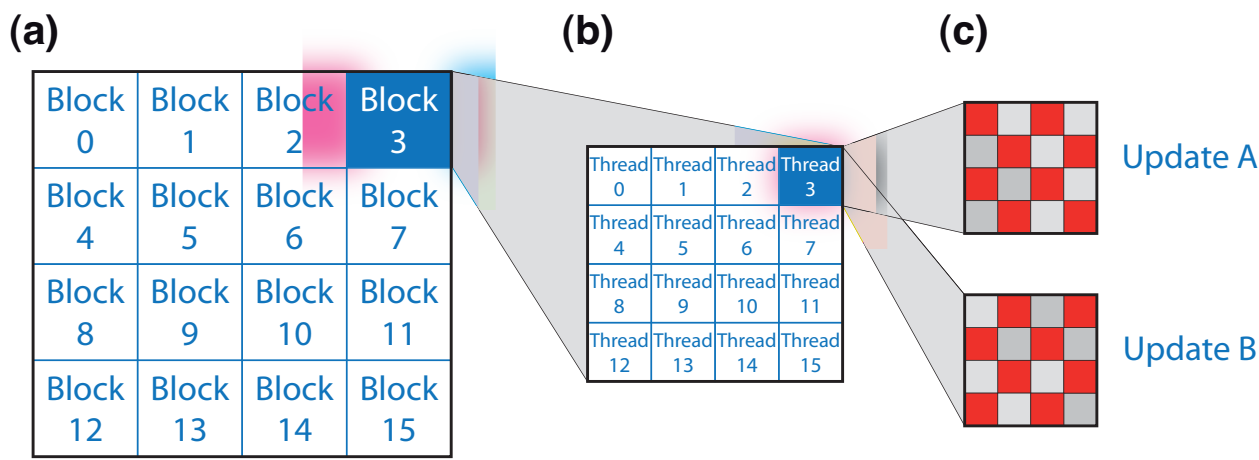}
\caption{(Color online) The spin lattice is processed by a variable number of blocks (a), where each block runs a variable number of threads (b). The threads update the spin lattice in two steps, $A$ and $B$, using two kernel invocations (c).}
\label{fig_newscheme}
\end{figure}

\begin{figure}[htb]
\includegraphics[width=\textwidth]{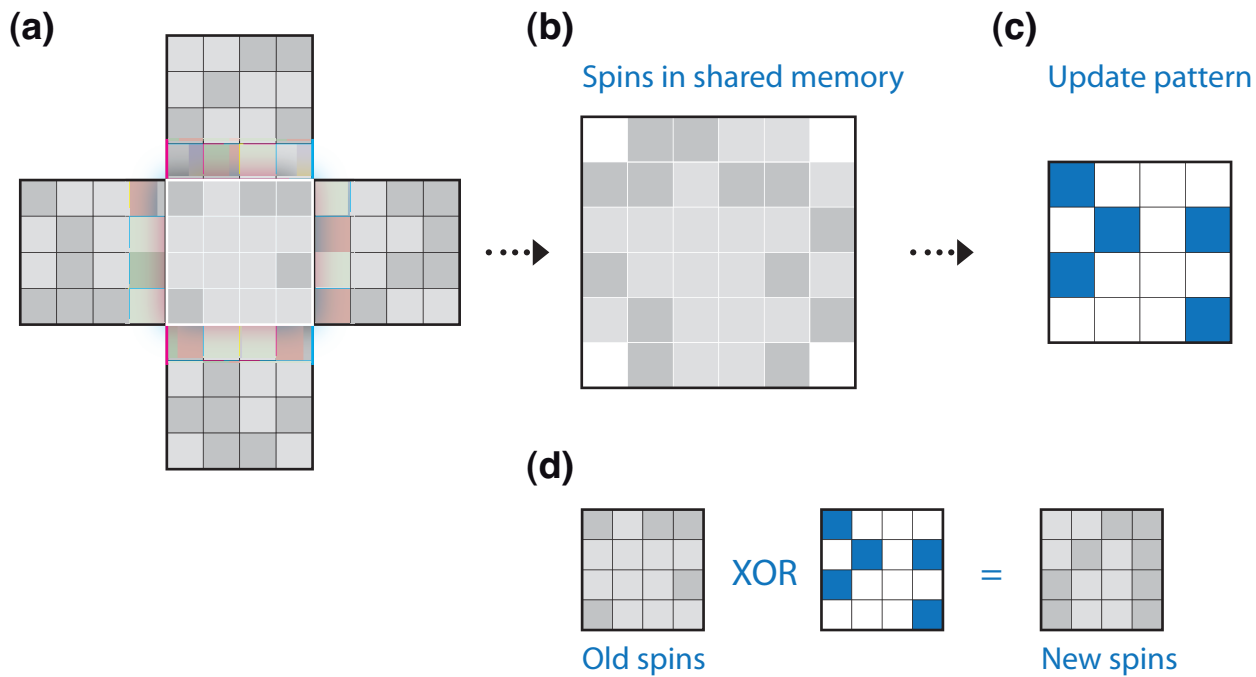}
\caption{(Color online) (a) The way a kernel processes a $4 \times 4$ meta-spin. (b) Spins are extracted into shared memory and an update pattern is created (c). (d) Afterwards, the new spins are obtained using the update pattern (Spins on blue sites will be flipped, spins on white sites will not be flipped), and written back to global memory.}
\label{fig_multi_update}
\end{figure}

The main goal of the following implementation is to reduce access to the global memory of the GPU, which is extremely costly.
The best performance without pre-calculating flip patterns can be achieved by extracting the single spins into shared memory and performing the calculations on integer registers. The spin field on the graphics card is encoded in quadratic blocks of $4\times 4$ spins (hereafter referred to as ``meta-spins'') which can be stored as binary digits of one unsigned short integer (2 bytes), which can be accessed by a single memory lookup. Single spin values can be extracted from one meta-spin for example using the expression
\begin{verbatim}
s[x,y] = ((meta-spin & (1 << (y * 4 + x)) != 0) * 2 - 1)
\end{verbatim}
which returns a value of either $-1$ or $1$. Here, {}``<<'' denotes a bitwise left-shift operation This is a slightly more complicated expression than for a linear layout, but it makes sense for a multi-GPU implementation, where border information has to be transferred between various GPUs (see Sect.~\ref{sec_multi_gpu}). This approach realizes that each spin uses exactly one bit of memory. The spin field is stored in global memory, which is expensive to access.

To process the spin field on the GPU, the spin field is subdivided into quadratic subfields which can be processed by threads grouped into one block (see Fig.~\ref{fig_newscheme}). Each thread of this block processes a ``meta-spin'' of $4\times 4$ spins. At the beginning of a kernel, it retrieves $5$ meta-spins from the global memory, namely its own and its four neighboring meta-spins (Fig.~\ref{fig_multi_update}a). This information is used to extract the information of the $4\times 4$ spins. Each thread will store the spin field of $4\times 4$ spins as well as the neighboring spins in a $6 \times 6$ integer array in shared memory, which allows for fast computation of the spin flips. The spin update is performed in two steps as described before. A first kernel is needed to update the ``black sites'' on a checkerboard pattern, and a second processes the ``white sites''. The update kernel for the white sites has to wait until all black sites have been updated. Thus, two separate kernels are needed. There is no other way to achieve global synchronization between the threads. 

Each kernel creates an update pattern, where each binary digit indicates if the associated spin has to be flipped or not. At the end of the kernel execution, the $4 \times 4$ meta-spins are updated with one single global memory write.

In summary, each update thread executes the following steps:

\begin{enumerate}
\item Look up meta-spins from global memory (See figure 2a)
\item Extract meta-spins into 6x6 integer array in shared memory which then contains the 4x4 meta-spins and the neighbors (See figure 2b)
\item For all 8 white/black sites $s_i$ in the 4x4 field, draw a random number and evaluate the Metropolis criterion
\item Generate the update pattern (set the $i$th bit to 1, if the flip of the $i$th was accepted) (Figure 2c)
\item Update the meta-spin by an XOR operation with the update pattern to obtain the spins at the next timestep (Figure 2d)
\end{enumerate}

Although the update scheme sounds hardly efficient, it dramatically reduces global memory access compared to the previous implementation, which results in faster computation times on GPU hardware.

After the update is completed, the magnetization per spin $m(T)$ has to be extracted from the lattice. In a first step, the magnetization of each block can be summed using the shared memory of each block by employing a binary tree reduction and writing out the total magnetization of the slice back to the main memory. The final summation of the magnetizations
of the blocks can be done either on the CPU or on the GPU at about equal speeds.

\subsection{Random number generation\label{sec_random_number}}

For every update thread a random number is needed, either to decide if the spin is flipped or not, or to look up an update pattern in global memory. This is why an efficient method to create random numbers is needed. 

In our implementation, we use an array of linear congruential random number generator (LCRNG) which is one of the oldest and most studied algorithm to generate pseudo random numbers \cite{Sch06}. A single random number generator provides the random numbers for every update thread $j$. A sequence of random numbers for the $j$th thread $x_{i,j}$ (where $i\in \mathbb{N}$) is generated by the recurrence relation 
\begin{equation}
\displaystyle x_{i+1,j}=(a\cdot x_{i,j}+c)\text{mod}\: m
\end{equation}
where $a,c$ and $m$ are integer coefficients. An appropriate choice of these coefficients is responsible for the quality of the produced random numbers. We use $a=1664525$ and $c=1013904223$ as suggested, e.g., in \cite{Pre07c}. Since by construction, results on a $32$-bit register are truncated to the endmost 32 bits, the modulo operation $m$ is set to $2^{32}$. By normalizing ($y_{i,j}=\text{abs}(x_{i,j}/2^{31})$) the LCRNG can be used to generate random numbers $y_{i,j}$ in the
interval $[0;1)$. For the GPU, an array of random numbers that provides a single random number seed for every spin update thread can be generated by the iteration
\begin{equation}
\displaystyle x_{0,j+1}=(16807\cdot x_{0,j})\text{ mod}\: m
\end{equation}
with $x_{0,0}=1$.

\subsection{Performance comparison}

\begin{figure}[htb]
\includegraphics[angle=-90,width=\textwidth]{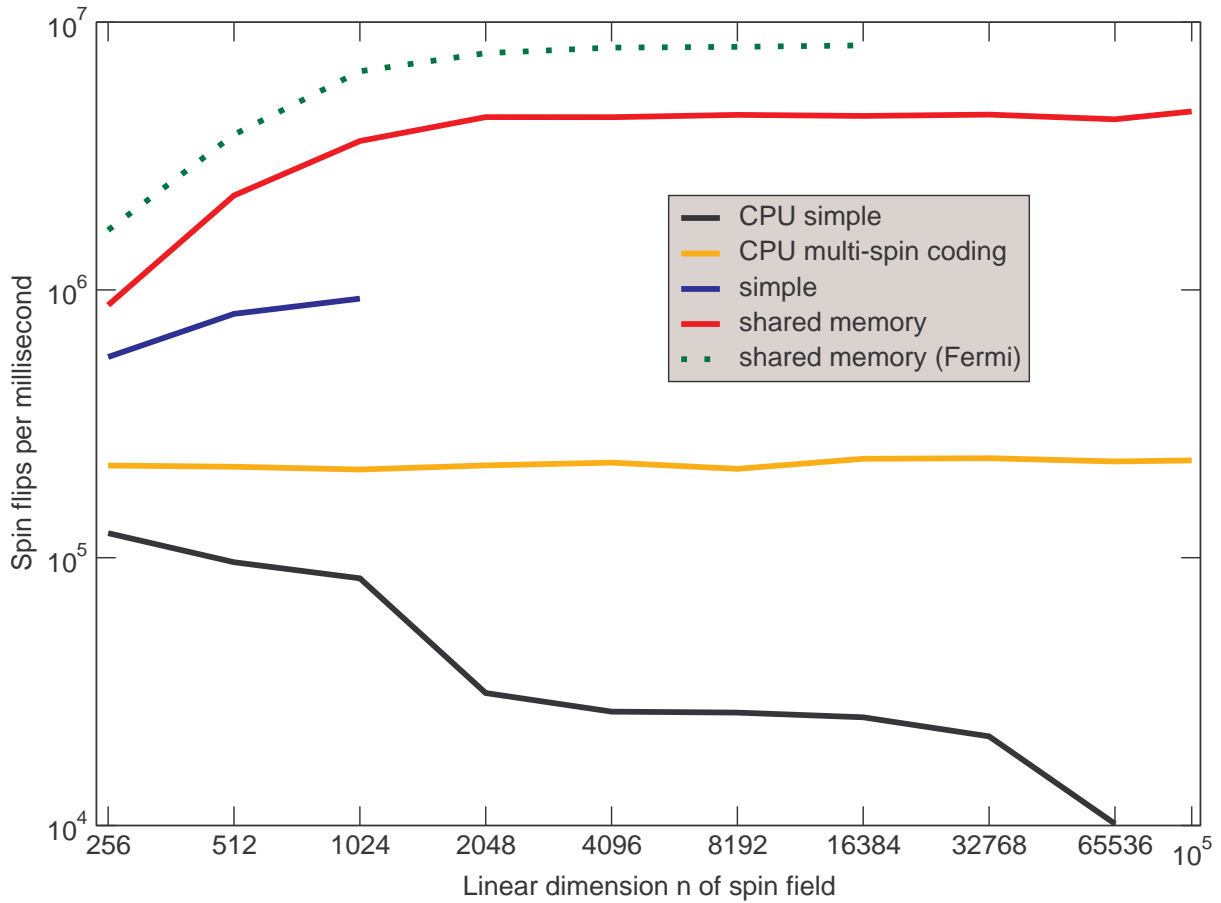}
\caption{(Color online) Benchmarking the implementations: The system is simulated at a constant temperature of $T= 0.99~T_{C}$. The performance of the straight port varies strongly with lattice size because of the large block size of $512$ threads, while the shared memory implementation offers stable performance over a wide range of sizes and offers better quality random numbers (comparable to the simple CPU implementation). The dotted line shows a preliminary benchmark of a Fermi GPU which became available in April 2010. A GeForce GTX 480 provides the following features: $1536$ MB global memory, $480$ streaming processor cores, $1.40$ GHz processor clock, $1848$ MHz memory clock, and a maximal power consumption of $250$ W.}
\label{fig_times}
\end{figure}

The multi-spin implementations are compared to simple implementations on both CPU and GPU. As a measurement for the performance of an implementation, we use the number of single spin flips per second, which also allows to compare results for different lattice sizes. The temperature is set to $0.99~T_C$.

The GPUs perform most efficient for lattice sizes of a linear dimension beyond $4096\times4096$. For this lattice size, a GPU is faster by a factor of about $15$--$35$, depending on the implementation and the resulting quality of random numbers. For the simple implementation used in \cite{Preis09}, between $1024\times1024$ and $2048\times2048$ spins the spin field size becomes comparable to the CPU L3 cache size, which leads to a higher rate of costly L3 cache misses. This is the point at which the simple implementation becomes inefficient.

%


\begin{table}
\caption[Comparison at lattice size $4096 \times 4096$]{Comparison at lattice size $4096 \times 4096$: \emph{CPU simple} encodes one spin in one integer, \emph{ CPU multi-spin coding} uses the efficient ``multi-spin'' update scheme presented in section~\ref{sec_cpu_implementation}, \emph{multi-spin unmodified} is a straightforward porting of this update to the GPU, \emph{multi-spin coding on the fly} uses the same scheme but calculates the update patterns at each update step on the fly, and \emph{multi-spin coding linear} determines one starting position in the random number pattern in the pool per block, and lets the threads read the random numbers linearly from that position on. The \emph{shared memory} implementation (see section~\ref{sec_shared_memory}) provides best quality random numbers. A preliminary test run on a Fermi GPU shows a factor of 1.82 compared to a Tesla C1060.}
\label{tab_comparing}
\begin{center}
\begin{tabular}{lrr}
\hline\noalign{\smallskip}
  & Spinflips per $\mu$s & Relative speed \\
\noalign{\smallskip}\hline\noalign{\smallskip}
\emph{CPU simple} & 26.6 & 0.11 \\
\emph{CPU multi-spin coding} & 226.7 & 1.00 \\
\emph{shared memory} & 4415.8 & 19.50 \\
\emph{shared memory (Fermi)} & 8038.2 & 35.46 \\
\emph{multi-spin unmodified}  & 3307.2 & 14.60 \\
\emph{multi-spin coding on the fly} & 5175.8 & 22.80 \\
\emph{multi-spin coding linear} & 7977.4 & 35.20 \\
\noalign{\smallskip}\hline
\end{tabular}
\end{center}
\end{table}

\section{Multi-GPU approach}\label{sec_multi_gpu}

\subsection{Implementation}

\begin{figure}[htb]
\includegraphics[width=\textwidth]{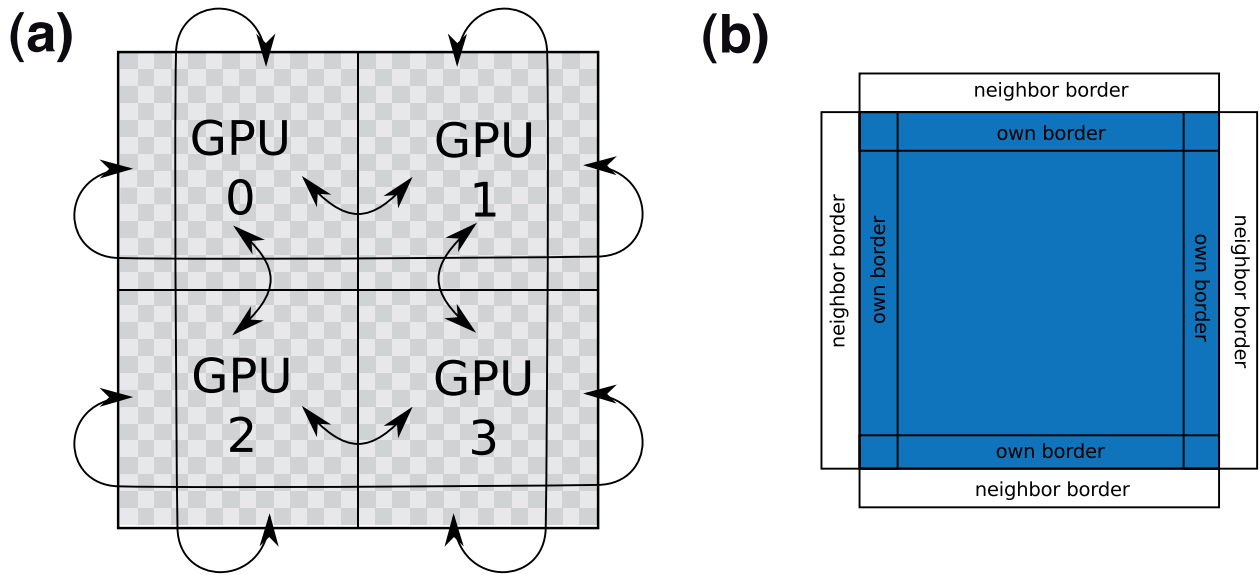}
\caption{(Color online) (a) Each GPU processes a ``meta-spin'' lattice of size $N=n^2$. The lattices are aligned on a super-lattice, and the outer borders are connected via periodic boundary conditions. In this example, $4$ GPUs process a system of $2^2 \cdot N$ spins. (b) A meta-spin update needs the 4 nearest neighbor meta-spins. On the borders of a lattice, each GPU needs the spin information of the neighboring lattices. The border information has to be passed between the GPUs. In our implementation this is done by using 8 neighbor arrays.}
\label{fig_borders}
\end{figure}

The general idea is to extend the quadratic lattice by putting multiple quadratic ``meta-spin'' lattices next to each other in a super-lattice (see Fig.~\ref{fig_borders}a for a 2$\times$2 super-lattice) and let each lattice be handled by one of the installed GPUs.  On the border of each lattice, at least one of the neighboring sites is located in the memory of another GPU (see Fig.~\ref{fig_borders}b). For this reason, the spins at the borders of each lattice have to be transferred from one GPU to the GPU handling the adjacent lattice. This can be realized by introducing four neighbor arrays containing the spins of the lattices' own borders, and four arrays for storing the spins of its adjacent neighbors (Fig.~\ref{fig_borders}c).

At the beginning of the execution, each MPI process initializes its own spin lattice, writes out its border spins into its own border arrays and sends them to its neighbors. In return it receives the adjacent borders from the according MPI processes. After this initialization phase, spins and random seeds are transferred to the GPU.

Then, a single lattice update has to be performed in the following way:

\begin{multicols}{2}
\begin{enumerate}
\item Copy neighbor borders to GPU memory
\item Call kernel to perform update (A)
\item Call kernel to extract borders from the spin array to own borders array
\item Copy own borders to host memory
\item Exchange borders with the other MPI processes
\item Copy neighbor borders to GPU memory again
\item Call kernel to perform update (B)
\item Call kernel to extract borders from spin array again
\item Transfer own borders to host memory
\item Exchange borders with other MPI processes
\item Retrieve processed data from GPU
\end{enumerate}
\end{multicols}

It turns out that the transfer time was not the limiting factor for our purposes but rather the latency of the memory accessed.

\subsection{GPU cluster performance}

\begin{figure}[htb]
\includegraphics[width=\textwidth]{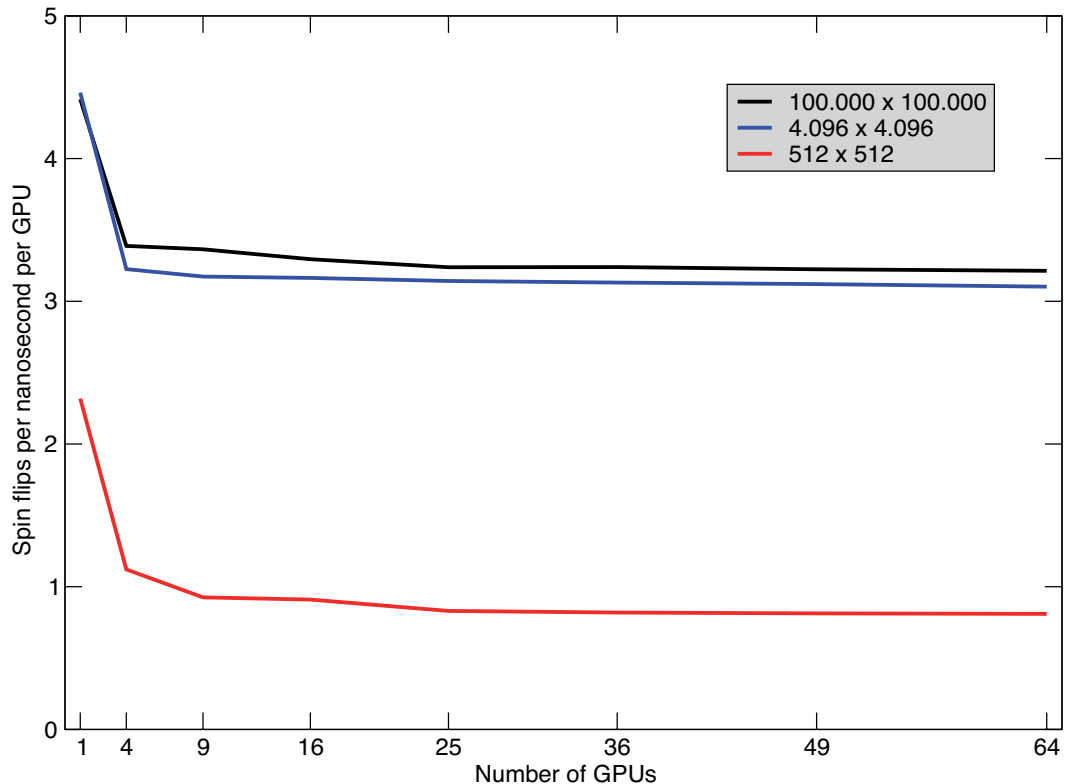}
\caption{(Color online) Cluster performance for various system sizes (per GPU). For more than one GPU, spin flip performance scales nearly linearly with the amount of GPUs. Again, optimal performance is reached at a lattice size of about $4096\times 4096$ per GPU. Using $64$ GPUs, a performance of $206$ spinflips per nanosecond can be achieved on a $800.000 \times 800.000$ lattice.}
\label{fig_times_cluster}
\end{figure}

For performance measurements on the GPU cluster, the shared memory implementation (see Sect.~\ref{sec_shared_memory}) was used, since it provided stable performance for various lattice sizes and because the memory layout is symmetric in the $x$ and $y$ direction, resulting in symmetric communication data. The tests were run on a GPU cluster with two Tesla C1060 GPUs in each node. Communication is established via Double Data Rate InfiniBand. The performance for various system sizes (see Fig.~\ref{fig_times_cluster}) provides evidence that for more than one GPU, spin flip performance scales nearly linearly with the amount of GPUs. The drop from one GPU to four GPUs is due to the communication overhead produced by exchanging borders. For larger system sizes, the communication overhead per CPU/GPU remains constant. An optimal performance is reached for lattice sizes beyond $4096\times 4096$ per GPU. For 64 GPUs---the NEC Nehalem Cluster maintained by the High Performance Computing Center Stuttgart (HLRS) provides $128$ GPUs---, a performance of $206$ spinflips per nanosecond can be achieved on a $800.000^2$ 2D Ising lattice, i.e. we can update the whole lattice in about three seconds.

\section{Conclusion}\label{sec_conclusion}

We presented two major improvements over our previous work. By using multi-spin coding techniques, we improved the computation to memory access ratio of our calculations dramatically, resulting in better overall performance. On a single GPU, up to 7.9 spinflips per nanosecond are possible,  $15$ to $35$ times faster than our highly optimized CPU version , depending on the implementation and the quality of random numbers. The other improvement targets the utilization of GPU clusters, where the 2D Ising lattice is distributed over many GPUs. We show that our implementation scales nearly linearly with the number of GPUs, which allows us to process huge Ising lattices on GPU clusters. 
Preliminary tests on a NVIDIA GPU of the latest generation---the Fermi architecture which offers twice the amount of streaming processor cores---indicate an additional speedup of roughly 1.8 compared to a Tesla C1060.

\section*{Acknowledgments}

We thank K.~Binder, D.~Stauffer, and M.~Tacke for fruitful discussions. GPU time was provided on the NEC Nehalem Cluster by the High Performance Computing Center Stuttgart (HLRS). We are grateful to T.~B{\"o}nisch, B.~Krischok, and H.~P{\"o}hlmann for their support at the HLRS. This work was financially supported by the Deutsche Forschungs\-gemein\-schaft (DFG) and benefited from the Gutenberg-Akademie and the Schwerpunkt f{\"u}r rechner\-ge\-st{\"u}tzte Forschungsmethoden in den Naturwissenschaften and the Materialwissenschaftliches Forschungszentrum of the Johannes Gutenberg University Mainz.


\end{document}